\documentclass[journal]{IEEEtran}

\usepackage{amsmath}
\usepackage{amssymb}
\usepackage{bm}
\usepackage{graphicx}
\usepackage{cite}
\usepackage{algorithmic}
\usepackage{algorithm}

\newcommand{\bH}{\ensuremath{\bm{H}}}

\newcommand{\br}{\ensuremath{\bm{r}}}
\newcommand{\bv}{\ensuremath{\bm{v}}}

\newcommand{\bp}{\ensuremath{\bm{p}}}
\newcommand{\bn}{\ensuremath{\bm{n}}}
\newcommand{\varn}{\ensuremath{\sigma_{n}^{2}}}
\newcommand{\bI}{\ensuremath{\bm{I}}}
\newcommand{\sV}{\ensuremath{\mathcal{V}}}
\newcommand{\bc}{\ensuremath{\bm{c}}}
\newcommand{\bC}{\ensuremath{\bm{C}}}
\newcommand{\bs}{\ensuremath{\bm{s}}}
\newcommand{\bu}{\ensuremath{\bm{u}}}

\newcommand{\bvh}{\ensuremath{\hat{\bm{v}}}}
\newcommand{\bvb}{\ensuremath{\bar{\bm{v}}}}

\newcommand{\bb}{\ensuremath{\bm{b}}}
\newcommand{\bS}{\ensuremath{\bm{S}}}

\newcommand{\bvt}{\ensuremath{\tilde{\bm{v}}}}

\newcommand{\B}{\ensuremath{\bar{B}}}
\newcommand{\K}{\ensuremath{\bar{K}}}
\newcommand{\diff}{\ensuremath{\mathrm{d}}}
\newcommand{\bvr}{\ensuremath{\bv_{(r)}}}
\newcommand{\bvi}{\ensuremath{\bv_{(i)}}}
\newcommand{\me}{\ensuremath{\mathrm{e}}}

\newtheorem{lemma}{Lemma}
\newtheorem{theorem}{Theorem}

\begin{document}

\title{Tree-Structured Random Vector Quantization for Limited-Feedback
  Wireless Channels}

\author{Wiroonsak Santipach~\IEEEmembership{Member,~IEEE,} and Kritsada Mamat%
\thanks{This work was supported by Thailand's Commission on Higher
  Education and the Thailand Research Fund under grant MRG5080174 and
  the 2008 Telecommunications Research and Industrial Development
  Institute (TRIDI) scholarship.  The material in this paper was
  presented in part at the ECTI conference, Krabi, Thailand, May 2008 and the
  IEEE International Conference on Telecommunications (ICT), Doha,
  Qatar, April 2010.}%
\thanks{The authors are with the Department of Electrical Engineering;
  Faculty of Engineering; Kasetsart University, Bangkok, 10900,
  Thailand (email: wiroonsak.s@ku.ac.th; g5317500192@ku.ac.th).}}

\markboth{IEEE Transactions on Wireless Communications}%
{Santipach and Mamat: Tree-Structured Random Vector Quantization for
Limited-Feedback Wireless Channels}

\maketitle

\begin{abstract}
We consider the quantization of a transmit beamforming vector in
multiantenna channels and of a signature vector in code division
multiple access (CDMA) systems.  Assuming perfect channel knowledge,
the receiver selects for a transmitter the vector that maximizes the
performance from a random vector quantization (RVQ) codebook, which
consists of independent isotropically distributed unit-norm
vectors. The quantized vector is then relayed to the transmitter via a
rate-limited feedback channel.  The RVQ codebook requires an
exhaustive search to locate the selected entry.  To reduce the search
complexity, we apply generalized Lloyd or $k$-dimensional (kd)-tree
algorithms to organize RVQ entries into a tree.  In examples shown,
the search complexity of tree-structured (TS) RVQ can be a few orders
of magnitude less than that of the unstructured RVQ for the same
performance.  We also derive the performance approximation for TS-RVQ
in a large system limit, which predicts the performance of a
moderate-size system very well.
\end{abstract}


\IEEEpeerreviewmaketitle

\section{Introduction}

Channel information at both the transmitter and receiver can increase
the performance of wireless systems significantly.  With channel
information, the transmitter can adapt its transmit power and waveform
to a dynamically fading channel while the receiver can detect
transmit symbols from received signals.  Typically, channel
information can be estimated at the receiver from pilot signals during
 a training period.  The transmitter on the other hand is usually not
able to directly estimate a forward channel, especially in a frequency
division duplex where a channel in one direction is not reciprocal to
that in the opposite direction.  Thus, the transmitter has to rely on
the receiver for channel information.  Normally, the receiver relays
channel information to the transmitter via a rate-limited feedback
channel.

The receiver can directly quantize channel coefficients and feeds back
the quantized coefficients to the transmitter, which adapts its
transmission, accordingly~\cite{commag04}.  Alternatively, the
receiver computes and quantizes the optimal transmit coefficients.
References~\cite{narula98,love03,mukkavilli03,roh_it06,mimo09}
proposed quantization of the transmit precoding matrix, which consists of
transmit antenna weights in a multiantenna channel while
\cite{cdma05,w_dai} considered quantization of the signature vector in
CDMA.  Comparing the two approaches, \cite{commag04} showed that
quantizing transmit coefficients performs much better than direct
channel quantization.

Most of the proposed quantization codebooks require an exhaustive
search to locate the quantized signature vector.  The search
complexity depends on the number of entries in the quantization
codebook, which grows exponentially with available feedback bits
denoted by $B$.  Reducing the search complexity while not jeopardizing
the performance is desirable.  In~\cite{cdma05}, the number of
coefficients to be quantized was reduced by projecting the signature
vector onto a lower dimensional subspace and the coefficients were
then scalar quantized.  Although the complexity of scalar quantization
is much less than that of vector quantization, the performance of the
scalar quantization scheme suffers
greatly. References~\cite{ryan09,McKilliam2008a} proposed a search
algorithm for the QAM codebook, which is based on a noncoherent
detection algorithm~\cite{glrt}.  The associated complexity grows
linearly with $B$ with small performance loss.  However, the algorithm
is not applicable when $B$ is smaller than the number of
coefficients. To reduce the search complexity, we propose to organize
an {\em unstructured} random vector quantization (RVQ) codebook into a
tree by either a generalized Lloyd (GLA)
algorithm~\cite{max60,lloyd82} or a $K$-dimensional (kd) tree
algorithm~\cite{bentley80,bentley90}.  The tree-building process is
computationally complex, but can be performed offline.  An RVQ
codebook was proposed in~\cite{mimo09,cdma05} and contained
independent isotropically distributed unit-norm vectors.  For
moderate-size systems, RVQ performs close to the optimum codebook
designed for a channel with independent identically distributed gains.
In a large system limit in which system parameters tend to infinity
with fixed ratios, RVQ is optimal~\cite{mimo09,cdma05} (i.e.,
maximizing capacity or minimizing interference power).

In this work, we consider quantizing both the transmit beamforming
vector in multi-input multi-output (MIMO) channel and the signature
sequence in reverse-link CDMA.  With a tree-structured (TS) RVQ
codebook, the receiver searches the tree for the vector that is
closest in Euclidean distance to the optimal {\em unquantized}
signature vector, which is the eigenvector of the received covariance
matrix.  We note that the obtained quantized vector may not give the
optimal performance (i.e, maximizing received power in a MIMO channel or
minimizing the interference power in CDMA).  However, the performance
loss incurred is small when $B$ is large, while the corresponding
search complexity is reduced by a few orders of magnitude from that of
a full search.  For smaller $B$, the performance, however, takes a
significant loss.  So, we modify the kd-tree search algorithm to
narrow the performance gap. Analyzing the performance of the TS-RVQ
codebook is not tractable.  Hence, we derive the performance
approximation of the TS-RVQ codebook in the large system limit.  The
derived approximation is a function of the number of feedback bits per
degree of freedom and the normalized load and is shown to predict the
performance of a system with moderate-to-large size very well.

\section{System Model}
\label{sys_mod}

We are interested in two wireless channel models as follows.

\subsection{Point-to-Point Multiantenna Channel}

We first consider a discrete-time point-to-point channel with $N_t$
transmit antennas and $N_r$ receive antennas.  The $N_r \times 1$
received vector is given by
\begin{equation}
  \br = \bH \bv b + \bn
\end{equation}  
where $\bH = [h_{n_r,n_t}]$ is an $N_r \times N_t$ channel matrix,
$\bv$ is an $N_t \times 1$ beamforming unit-norm vector, $b$ is a
transmitted symbol, $\bn$ is an $N_r \times 1$ additive white Gaussian
noise vector with zero mean and covariance $\varn \bI$, and $\bI$ is
an identity matrix.  Assuming an ideal rich scattering environment, we
model $h_{n_r,n_t}$ as a complex Gaussian random variable with zero
mean and variance $1/N_r$. Thus, the received fading power is given by
$\sum_{n_r = 1}^{N_r} E [ | h_{n_r,n_t} |^2] = 1$ for a given $n_t$th
transmit antenna.  An ergodic channel capacity is given by
\begin{equation}
 C(\bv) = E_{\bH} [ \log ( 1 + \rho \bv^{\dag} \bH^{\dag} \bH \bv ) ]
 \label{capacity}
\end{equation}
where expectation is over distribution of $\bH$, and $\rho = 1/\varn$
is the background signal-to-noise ratio (SNR).  We note that the
capacity \eqref{capacity} is a function of the beamforming vector $\bv$
and that only a rank-one transmit beamforming is considered.
Quantization of an arbitrary-rank transmit precoding matrix was
studied by~\cite{mimo09}.

\subsection{Reverse-Link CDMA}

We also examine a reverse-link CDMA with $K$ users and processing gain
$N$.  User $k$ is assigned the $N \times 1$ signature vector $\bs_k$
for $1 \le k \le K$. The user signal is assumed to traverse $L$ fading
paths.  We denote the $N \times N$ channel matrix for user $k$ by
\begin{equation}
  \bC_k = \left[\begin{array}{ccccccc}
      h_{k,1}&0&\ldots&0&0&\ldots&0 \\\vdots&h_{k,1}& &\vdots&0&
      &\vdots \\h_{k,L}&\vdots&\ddots&0&\vdots& &0 \\0&h_{k,L}&
      &h_{k,1}&0&\ldots&0 \\\vdots&0&\ddots&\vdots&h_{k,1}& &0
      \\0&\vdots& &h_{k,L}&\vdots&\ddots&0
      \\0&0&\ldots&0&h_{k,L}&\ldots&h_{k,1}
  \end{array}\right]
\end{equation}
where fading gains for user $k$, $h_{k,1}$, \ldots, $h_{k,L}$ are
independent complex Gaussian random variables with zero mean and
variance $E|h_{k,1}|^2$, \ldots, $E|h_{k,L}|^2$, respectively.  Here
we assume that the symbol duration is much longer than the delay spread
and thus, inter-symbol interference is negligible.  For an ideal
nonfading channel, $\bC_k$ = $\bI$ for all $k$.  

The $N \times 1$ received vector at the base station is given by
\begin{equation}
  \br = \bH \bb + \bn
\end{equation}
where $\bb = [b_1 \quad b_2 \ \ldots \ b_K]$ is the vector of $K$
users' transmitted symbols and
\begin{equation}
  \bH = [ \bC_1 \bs_1, \ \ldots \ , \bC_K \bs_K] .
\end{equation}

In this work, we only consider single-user signature quantization.
Without loss of generality, we assume that user 1's signature is
quantized while other signatures do not change.  Signature
quantization for multiple users was considered by~\cite{ts09}.
Assuming a matched-filter receiver for user 1
\begin{equation}
  \bc_1 = \frac{\bC_1\bs_1}{|\bC_1\bs_1|},
\end{equation}
the interference power for user 1 is given by
\begin{equation}
  I_1 \triangleq \sum_{k=2}^K(\bs_1^\dag \bC_1^\dag\bC_k\bs_k)^2 =
  \bs_1^\dag \bC_1^\dag \bH_1^{\dag}\bH_1\bC_1 \bs_1
\end{equation}
where $\bH_1 = [ \bC_2 \bs_2 \quad \bC_3 \bs_3 \ \ldots \ \bC_K
  \bs_K]$ is the $N \times (K-1)$ interfering signature matrix.
Hence, the associated output signal-to-interference plus noise ratio
(SINR) for user 1 is given by
\begin{equation}
  \gamma_1(\bs_1) = \frac{(\bs_1^\dag\bC_1^\dag\bC_1\bs_1)^2}{I_1 + \varn
    \bs_1^\dag\bC_1^\dag \bC_1\bs_1} .
\label{sinr_chan}
\end{equation}
Similar to MIMO channels, the performance of the CDMA user
\eqref{sinr_chan} also depends on the transmit signature vector
$\bs_1$.

With channel information, the receiver in the MIMO channel selects $\bv$
that maximizes the capacity from an RVQ codebook, which is known {\em
  a priori} at both the transmitter and receiver.  The RVQ codebook
contains independent and isotropically distributed unit-norm vectors
and is denoted by
\begin{equation}
  \sV = \{\bv_1,\bv_2,\ldots,\bv_{2^B}\}
\label{rvq_cb}
\end{equation}
where $B$ denotes available feedback bits.  RVQ was proposed
by~\cite{cdma05,mimo09} and was shown to be optimal in a large system
limit in which $(N_t,N_r,B) \to \infty$.  For CDMA, the receiver
selects $\bs_1$ that minimizes the interference power for user $1$.

Maximizing the capacity for the MIMO channel in \eqref{capacity} is
equivalent to maximizing the received signal power.  Thus, the
receiver selects
\begin{equation}
  \bvh = \arg \max_{\bv_j \in \sV} \ \bv_j^{\dag}\bH^{\dag} \bH \bv_j
\label{bvh}
\end{equation}
for given $\bH$.  In CDMA, the receiver selects
\begin{equation}
  \hat{\bs}_1 = \arg \min_{\bv_j \in \sV} \ \bv_j^{\dag}\bC_1\bH_1^{\dag}
  \bH_1 \bC_1^{\dag}\bv_j
\label{bs1}
\end{equation}
for given $\bH_1$ and $\bC_1$ to minimize the interference power.  As
$B$ increases, the number of entries in the codebook increases and so
does the performance with the selected signature vector in both
\eqref{bvh} and \eqref{bs1}.  However, the search complexity also
grows since the RVQ codebook requires a full search to locate the
optimal entry and thus, complexity can pose a serious problem for very
large $B$.

\section{Nearest-Neighbor Search}
\label{tsrvqcb}

Let $\bS\bS^{\dag}$ denote a covariance matrix where $\bS$ denotes an
$N \times K$ effective signature matrix for both MIMO and CDMA
channels.\footnote{For MIMO, $\bS\bS^{\dag} = \bH^{\dag} \bH$ while
  for CDMA, $\bS\bS^{\dag} = \bC_1^\dag \bH_1^{\dag}\bH_1\bC_1$.}
Performing a singular value decomposition (SVD) on the covariance
matrix gives
\begin{equation}
  \bS \bS^{\dag} = \sum_{i = 1}^N \lambda_i \bu_i \bu^{\dag}_i
\label{SS}
\end{equation}
where $\lambda_1 \le \lambda_2 \le \cdots \le \lambda_N$ are the
ordered eigenvalues and $\bu_i$ is the corresponding $i$th
eigenvector.  It is well known that the first  eigenvector
maximizes the quadratic form $\bv^{\dag} \bS \bS^{\dag} \bv$ as follows 
\begin{equation}
  \bu_1 = \arg \max_{\bv} \bv^{\dag} \bS \bS^{\dag} \bv
\end{equation}
while the $N$th eigenvector minimizes the quadratic form as follows
\begin{equation}
\bu_N = \arg \min_{\bv} \bv^{\dag} \bS \bS^{\dag} \bv .
\end{equation}
To quantize these optimal eigenvectors for the transmitter, the
receiver may search the given codebook for the vector that is closest
in Euclidean distance to either $\bu_1$ or $\bu_N$ (finding the
nearest neighbor).  To quantize $\bu_1$, the receiver selects
\begin{equation}
  \bvb_1 \triangleq \arg \min_{\bv_j \in \sV} \|\bu_1 - \bv_j\|^2 = \arg
  \max_{\bv_j \in \sV} \Re\{ \bu^{\dag}_1 \bv_j \}
\label{bvb1}
\end{equation}
where $\Re\{z\}$ is the real part of $z$.  The right-hand side of
\eqref{bvb1} follows since $\|\bv_j\| = \|\bu_1\| = 1$. We note that
$\bvb_1$ needs to be updated whenever the channel covariance
$\bS\bS^{\dag}$ and consequently, its eigenvectors are changed.  The
associated performance is given by
\begin{equation}
  \bar{I}_{\max} \triangleq \bvb^{\dag}_1 \bS \bS^{\dag} \bvb_1 .
\end{equation}
Similarly, the receiver can also quantize $\bu_N$ by selecting
\begin{equation}
  \bvb_N \triangleq \arg \max_{\bv_j \in \sV}  \Re\{ \bu^{\dag}_N \bv_j \}
\end{equation}
with associated performance 
\begin{equation}
  \bar{I}_{\min} \triangleq \bvb^{\dag}_N \bS \bS^{\dag} \bvb_N .
\end{equation}

Finding the nearest neighbor is a classical vector quantization
problem to which there are many solutions~\cite{gersho} (see
references therein).  We note that $\bvb$ is suboptimal and may not
necessarily maximizes or minimizes the quadratic form (the received
signal power in \eqref{bvh} or the interference power in \eqref{bs1}).
However, the performance difference is minimal with a large codebook
(large $B$).  To avoid an exhaustive search to find the nearest
neighbor, we propose to organize the RVQ codebook entries into a tree by
applying either generalized Lloyd or kd-tree algorithms.
 
\subsection{Generalized Lloyd Algorithm}

We start with RVQ codebook $\sV$ with $2^B$ entries.  To build a
binary tree, we iteratively divide the RVQ entries into two groups with
the generalized Lloyd algorithm (GLA)~\cite{gersho}.  We note that with
this algorithm, the tree produced may not be balanced.  Building the
tree becomes more complex as $B$ increases.  However, it should not
incur any additional delay since the tree can be produced offline.

To find the optimal vector $\bvb$, we apply the encoding method
proposed by~\cite{tsnn96}.  Since the method is only applicable to
a real vector, we transform an $N \times 1$ complex vector $\bv_j$ into
a $2N \times 1$ real vector $[\bv_{j,r}^T \quad \bv_{j,i}^T]^T$, where
$\bv_{j,r} = \Re\{\bv_j\}$, $\bv_{j,i} = \Im\{\bv_j\}$, and $\Im\{z\}$
is the imaginary part of $z$.  The stated encoding method produces
$\bvb$ that is closest in Euclidean distance to the desired
eigenvector with search complexity growing approximately {\em
  linearly} with $B$.

\subsection{Kd-Tree Algorithm}

A kd-tree is a data structure for storing points in a $k$-dimensional
space.  The kd-tree algorithm produces an unbalanced binary-search
tree by clustering the codebook entries by dimension at each
step~\cite{bentley80,bentley90}.  Since the kd-tree algorithm searches for
the nearest neighbor of a real vector, we again transform an $N \times
1$ complex eigenvector into a $2N \times 1$ real vector before
quantization.  First, we construct an RVQ codebook with $2N \times 1$
real unit-norm vectors and store the codebook at the root node.  Then,
we calculate a median for the first elements of all vectors in the
codebook and select the element closest to the median as the {\em pivot}.
By comparing the first element of the vector to the pivot, all vectors
are divided into 2 groups, which will be stored in the left-and
right-child nodes.  Next we move to either the left- or right-child nodes
and compute the pivot for the second dimension to divide the vectors
into 2 groups for that node.  We operate on the next dimension each
time we move down the tree and iterate the process until each node
contains only one vector. The detailed steps for building kd-tree are
shown in~\cite{bentley80,bentley90}.  Building the kd-tree is
relatively faster than building the tree by GLA since we examine one
dimension at a time and this can be done offline.

To quantize the desired vector, we start at the root node.  By
comparing the element of the vector with the pivot of the present node
in the specified dimension, we move down to either the left- or
right-child nodes.  We transverse the tree until the leaf node is reached
and the candidate entry is obtained.  The algorithm proposed
in~\cite{bentley80,bentley90} makes certain that the candidate is the
nearest neighbor by comparing the distance of the candidate and the
vector to quantize with that of other nodes.  The kd-tree search produces
$\bvb$ that is closest in Euclidean distance to the desired
eigenvector among the entries in the RVQ codebook.  The associated search
complexity is a fraction of that of a full search as simulation
results will demonstrate.

\subsection{Performance Approximation}

Evaluating the quadratic form with the nearest-neighbor quantized
vector is an open and difficult problem.  Hence, we approximate the
nearest-neighbor criterion with the closest-in-angle one. For a
closest-in-angle search, the receiver selects
\begin{equation}
  \bvt_1 = \arg \max_{\bv_j \in \sV} \ |\bu^{\dag}_1 \bv_j|^2 = \arg
  \max_{\bv_j \in \sV} \ \cos^2 \phi_j
\label{eq_bvt1}
\end{equation}
where $\phi_j$ is the angle between $\bu_1$ and $\bv_j$, and the
corresponding performance is given by
\begin{align}
\tilde{I}_{\max} & \triangleq \bvt^{\dag}_1 \bS \bS^{\dag} \bvt_1 \label{eq_t1}\\
                & = \lambda_1 |\bvt^{\dag}_1 \bu_1|^2 + \sum_{i =
    2}^N \lambda_i |\bvt^{\dag}_1 \bu_i|^2  \label{bui2}
\end{align}
where SVD in \eqref{SS} is applied. (Similarly to \eqref{eq_bvt1} and
\eqref{eq_t1}, we can also define $\bvt_N$ and $\tilde{I}_{\min}$.)

Evaluating the expectation for $\tilde{I}_{\max}$ for finite $N$, $K$, and
$B$ is not tractable.  Hence we analyze $\tilde{I}_{\max}$ in a large
system limit where $(N, K, B) \to \infty$ with fixed ratios.  The
first term on the right-hand side in~\eqref{bui2} was shown
by~\cite{mimo09} to converge as follows
\begin{equation}
  |\bvt^{\dag}_1 \bu_1|^2 \to 1 - 2^{-\B}
\label{12b}
\end{equation}
almost surely as $(N, B) \to \infty$ with fixed $\B \triangleq B/N$.
With infinite feedback ($\B = \infty$), $\bvt_1 \to \bu_1$ and hence,
$|\bvt^{\dag}_1 \bu_1|^2 \to 1$.  On the other hand, with no feedback,
$\bvt_1$ is randomly selected and $|\bvt^{\dag}_1 \bu_1|^2 \to 0$.
The second term in~\eqref{bui2} converges to the following limit.
\begin{lemma}
\label{l1}
  As $(N, B, K) \to \infty$ with fixed $\B = B/N$ and $\K = K/N$, we
  have
  \begin{equation}
    \sum_{i=2}^N\lambda_i |\bvt_1^{\dag}\bu_i|^2 \to 2^{-\B}
    \int_0^{\infty} \lambda g_{\bS\bS^{\dag}}(\lambda) \, \diff
    \lambda
    \label{bdu}
  \end{equation} 
assuming that the eigenvalue density for $\bS\bS^{\dag}$ converges to
a deterministic function $g_{\bS\bS^{\dag}} ( \cdot )$.
\end{lemma}

\begin{IEEEproof}
Since entries in the RVQ codebook are uniformly distributed on the
unit-norm hypersphere, it was shown by~\cite{mukkavilli03} that $\Pr\{
|\bvt^{\dag}\bu_1|^2 \ge s_1^2 \}$ is proportional to the surface area
of the spherical cap of the $N$-dimensional hypersphere described by
\begin{equation}
   \| \bu_1 \|^2 =  \sum_{n = 1}^N |u_n|^2 = r
\end{equation}
intersecting with $| u_1 |^2 \ge s_1^2$, where $u_n$ is the $n$th
complex entry of $\bu_1$ and $r$ denotes the radius.\footnote{Later we will set
$r=1$.} The surface area of the described spherical cap is given
by~\cite{mukkavilli03}
\begin{equation}
  A_N(r, s_1) = \frac{2 \pi^N}{(N-1)\!} r(r^2 - s_1^2)^{N-1} .
\end{equation}

We would like to evaluate $\Pr\{|\bvt^{\dag}\bu_2|^2 \ge s_2^2 \}$
conditioned that $(\bvt^{\dag}\bu_1)^2 \ge s^2_1 $.  Similar to the 
results in~\cite{mukkavilli03}, we can deduce that the conditional
probability is proportional to the surface area of the intersection
between the two spherical caps.  The volume of the intersection is
given by
\begin{equation}
  \left\{ \bu_1 \in \mathbb{C}^N \mid u_n = r_n \me^{j \theta_n},
  \sum_{n = 1}^N r_n^2 = r, \ r_1^2 \ge s^2_1 \ \text{and} \ r_2^2 \ge
  s_2^2 \right\}
\end{equation}
and can be computed with spherical coordinates as follows
\begin{multline}
  V_N (r, s_1, s_2) = (2\pi)^N \int_{r_1 = s_1}^{\sqrt{1 - s^2_2}} \int_{r_2 =
    s_2}^{\sqrt{1 - r_1^2}} \\ \left( \idotsint\limits_{\sum_{n =
      3}^N r_n^2 \le r - r_1^2 - r_2^2} r_3 \cdots r_n \, \diff r_3
  \cdots \diff r_N \right) r_1 r_2 \, \diff r_2 \diff r_1 .
\label{vaa}
\end{multline}
We note that the multiple integral in the brackets in \eqref{vaa} is
the volume of an $(N-2)$-dimensional hypersphere with scaling factor
of $(2 \pi)^{N-2}$~\cite{mukkavilli03}.  Thus,
\begin{align}
   V_N (r, s_1, s_2) & = \frac{(2 \pi)^N}{2^{N-2} (N-2)\!} \int_{r_1 =
     s_1}^{\sqrt{1 - s^2_2}} \int_{r_2 = s_2}^{\sqrt{1 -
       r_1^2}} r_1 r_2 \nonumber \\
    &\quad  \times  (r - r_1^2 - r_2^2)^{N-2} \, \diff r_2 \diff
   r_1 \\
    & = \frac{(2 \pi)^N}{2^{N} N \!} (r - s^2_1 - s^2_2)^N .
\end{align}
The associated surface area is obtained by differentiating the volume
$V_N (r, s^2_1, s^2_2)$ with respect to $r$ and is given by
\begin{align}
  A_N(r, s_1, s_2) &= \frac{\diff V_N (r, s_1, s_2)}{\diff r} \\
                   &= \frac{2\pi^N}{(N-1)\!} r(r^2 - s_1^2 - s_2^2)^{N-1} .
\end{align}
Thus, the conditional probability
\begin{align}
  \Pr \{|\bvt^{\dag}\bu_2|^2 \ge s_2^2 & \mid |\bvt^{\dag}\bu_1|^2 \ge
  s^2_1 \} \nonumber \\
   & = \frac{\Pr\{|\bvt^{\dag}\bu_2|^2 \ge s_2^2,
    |\bvt^{\dag}\bu_1|^2 \ge s^2_1 \}}{\Pr\{|\bvt^{\dag}\bu_1|^2 \ge
    s^2_1 \}} \\
   & =  \frac{A_N(r, s_1, s_2)}{A_N(r, s_1)} \\
   & = \left(1 -
  \frac{s_2^2}{r^2 - s_1^2} \right)^{N-1}
\end{align}
and the corresponding cumulative distribution function (cdf) for
$|\bvt_1^{\dag}\bu_2|^2$ given that $|\bvt_1^{\dag}\bu_1|^2 \ge
s^2_1$ and $r = 1$ is given by
\begin{equation}
  F_{|\bvt_1^{\dag}\bu_2|^2}(x \mid |\bvt_1^{\dag}\bu_1|^2 \ge s^2_1) = 1
  - \left( 1 - \frac{x}{1 - s_1^2} \right)^{N-1} .
\label{lf1}
\end{equation}
For the $i$th eigenvector where $i \ne 1$, the cdf for
$|\bvt_1^{\dag}\bu_i|^2$ is the same as shown in~\eqref{lf1}.  Thus,
with the cdf, we can compute the conditional expectation as follows
\begin{equation}
  E[|\bvt_1^{\dag}\bu_i|^2 \mid |\bvt_1^{\dag}\bu_1|^2 \ge s^2_1] =
  \frac{1}{N} (1 - s_1^2)
\label{bvd}
\end{equation}
which is converging to zero as $N \to \infty$.  $\Pr\{
|\bvt_1^{\dag}\bu_1|^2 \ge s^2_1 \}$ depends on the number of entries in
the RVQ codebook or available feedback bits.  As $(N, B) \to \infty$,
\cite{mimo09} has shown that
\begin{equation}
  1 - s_1^2 \to 2^{-\B} .
\end{equation}
If the eigenvalue density of $\bS\bS^{\dag}$ converges to a
deterministic function $g_{\bS\bS^{\dag}}(\cdot)$,
\begin{equation}
  \frac{1}{N} \sum_{i=2}^N \lambda_i \to  \int_0^\infty \lambda
  g_{\bS\bS^{\dag}}(\lambda) \, \diff \lambda .
\label{f1Ns}
\end{equation}
as $(N,K) \to \infty$ with fixed $K/N$.  Combining~\eqref{bvd} and
\eqref{f1Ns}, we have Lemma~\ref{l1}.
\end{IEEEproof}

From Lemma~\ref{l1}, as $\B$ increases, $\bvt_1$ and $\bu_i$ are
becoming perpendicular and $|\bvt_1^{\dag}\bu_i|^2 \to 0$.  We apply
Lemma~\ref{l1} to obtain the following performance approximations for the
nearest-neighbor search for MIMO and CDMA models in
section~\ref{sys_mod}.

\begin{theorem}
\label{t1}
   As $(N_r, N_t, B) \to \infty$ with fixed $B/N_t$ and $\bar{N}_r =
   N_r/N_t$, the capacity of MIMO channel with the nearest-neighbor
   beamforming vector can be approximated as follows.
   \begin{align}
     C(\bvb_1) &\approx C(\bvt_1)\\
               &\to \log \Big( 1 + \rho\left(1 +
     \frac{1}{\sqrt{\bar{N}_r}}\right)^2 (1 - 2^{-\frac{B}{N_t}}) \nonumber\\
     &\qquad \quad +\frac{\rho}{\bar{N}_r} 2^{-\frac{B}{N_t}} \Big) .
   \end{align}
 \end{theorem}
\begin{IEEEproof}
  We evaluate the received power $\bvt^{\dag}_1 \bH^{\dag}\bH \bvt_1$
  by applying Lemma~\ref{l1} and~\eqref{12b}. Thus,
  \begin{multline}
    \bvt^{\dag}_1 \bH^{\dag}\bH \bvt_1 \to
    \lambda_{\max}^{\infty}(\bH^{\dag}\bH) (1 - 2^{-\frac{B}{N_t}}) \\+
    2^{-\frac{B}{N_t}} \int \lambda g_{\bH^{\dag}\bH}(\lambda) \,
    \diff \lambda
   \label{21f}
  \end{multline}
  $\bH^{\dag}\bH$ is a Wishart matrix with the well known asymptotic
  eigenvalue distribution.  In the large system limit, the maximum
  eigenvalue converges to~\cite{tulino_verdu}
  \begin{equation}
    \lambda_{\max}^{\infty}(\bH^{\dag}\bH) = \left(1 +
    \frac{1}{\sqrt{\bar{N}_r}}\right)^2 
  \end{equation}
  and
  \begin{equation}
    \int \lambda g_{\bH^{\dag}\bH}(\lambda) \, \diff \lambda =
    \frac{1}{\bar{N}_r}
  \label{1bN}
  \end{equation}
where $g_{\bH^{\dag}\bH}(\lambda)$ is given by~\cite{tulino_verdu}.
Combining the capacity expression in~\eqref{capacity} and
\eqref{21f}-\eqref{1bN} gives Theorem~\ref{t1}.
\end{IEEEproof}

\begin{theorem}
\label{thm_cdma}
  For CDMA, suppose the channel gain for user 1, $\sum_{l=1}^L
  E|h_{1,l}|^2 \to \alpha$ while channel gains for interfering users,
  $\sum_{k=1}^L E|h_{k,l}|^2 \to 1$ for $2 \le k \le K$ as $(N,K) \to
  \infty$.  The large system SINR for user 1 with the nearest-neighbor
  quantized signature and a matched filter is approximated as follows.
  \begin{align}
     \gamma(\bvb) &\approx \gamma(\bvt) \\
           &\to \left\{
  \begin{array}{l@{\quad:\ }l}
     \frac{\alpha}{\K 2^{1-\B} + \alpha\varn} & 0 \le \K \le
     1\\ \frac{\alpha}{(1-\frac{1}{\sqrt{\K}})^2(1-2^{-\B}) + \K
       2^{-\B} + \alpha\varn} & \K > 1
  \end{array} \right. .
   \end{align}
\end{theorem}
The proof is similar to that for Theorem~\ref{t1} with the asymptotic
minimum eigenvalue and eigenvalue density for the interference covariance
$\bC_1 \bH_1^{\dag} \bH_1\bC_1^{\dag}$ shown
in~\cite{tcom11,tulino_verdu}.  If the channel gains across
interfering users are not uniform (i.e., $\sum_{k=1}^L E|h_{k,l}|^2$
is arbitrary for $k$), we apply the asymptotic eigenvalue density of
$\bC_1\bH_1^{\dag} \bH_1 \bC_1^{\dag}$ with a non-uniform power
allocation from \cite{tcom11,tulino_verdu}.  In section~\ref{num_re},
we will compare the asymptotic approximations derived here with
simulation results.

\section{Modified Kd-Tree Search}

We modify the search for the kd-tree so that we transverse in the direction
that maximizes or minimizes the quadratic form $\bv^{\dag} \bS
\bS^{\dag} \bv$.  Suppose $\bvr = \Re\{\bv\}$ and $\bvi =
\Im\{\bv\}$.  Expanding the quadratic form gives
\begin{multline}
 \bv^{\dag} \bS \bS^{\dag} \bv = \bvr^{\dag}\Re\{ \bS \bS^{\dag} \}
 \bvr + \bvi^{\dag}\Re\{ \bS \bS^{\dag} \} \bvi \\ + \bvi^{\dag}\Im\{ \bS
 \bS^{\dag} \} \bvr + \bvr^{\dag}\Im\{ \bS \bS^{\dag} \} \bvi
\label{crs}
\end{multline}
where we use the fact that $\bS \bS^{\dag}$ is Hermitian.  We note
that the two cross terms in \eqref{crs} are much smaller than the first
two quadratic terms.  Hence, our objective is to maximize or minimize
the two quadratic terms in \eqref{crs}, which only depend on $\Re\{
\bS \bS^{\dag} \}$ .  To quantize $\bvr$, we start with the RVQ codebook
whose entry has norm $1/2$.  Similar to the nearest-neighbor search,
we start at the first dimension and determine the pivot.  Then, we
divide all entries into 2 child nodes and compute the second pivot for
each node.  We move to the child node whose pivot gives the larger
value of $\bvr^{\dag}\Re\{ \bS \bS^{\dag} \} \bvr$.  This differs from
the nearest-neighbor search, which does not take $\Re\{ \bS \bS^{\dag}
\}$ into account.  We continue the process until we reach the leaf
node and hence, produce the candidate vector.  The steps are
summarized in Algorithm~\ref{al9} and Algorithm~\ref{al10}, which
compares the candidate with the surrounding nodes.
\begin{algorithm}
\caption{Modified kd-tree search}
\begin{algorithmic}[1]

\STATE Start at the root of a kd-tree codebook with the channel covariance
matrix $\bS \bS^{\dag}$.

\STATE Set $n = 1$ where $n$ is the index of element in an
$N$-dimensional vector.

\WHILE{Not a leaf node}
      \IF{$n > N$}
         \STATE $n = 1$
      \ENDIF

      \STATE $\bp_L^{(n)} = [0 \quad 0 \ \ldots \ p_{L,n+1} \ \ldots
        \ 0 \quad 0]^\dag$ where $\bp_L$ is the vector of the
      left-child node.
      
      \STATE $\bp_R^{(n)} = [0 \quad 0 \ \ldots \ p_{R,n+1} \ \ldots
        \ 0 \quad 0]^\dag$ where $\bp_R$ is the vector of the
      right-child node.
      
      \IF{${\bp_L^{(n)}}^\dag\bS \bS^\dag\bp_L^{(n)} >
        {\bp_R^{(n)}}^\dag\bS\bS^\dag\bp_R^{(n)}$} 
           \STATE Move to the left-child node.  
      \ELSE 
           \STATE Move to the right-child node.
      \ENDIF 

      \STATE $n = n + 1$ 
\ENDWHILE

\STATE $\bv$ = vector from the leaf node. 

\STATE $I = \bv^\dag \bS \bS^\dag \bv$

\STATE $n = 1$ 

\STATE $\bs$ = compare(root node, $n$, $\bS$, $I$, $\bv$) 

\RETURN $\bs$
\end{algorithmic}
\label{al9}
\end{algorithm}

\begin{algorithm}
\caption{function: $\bs$ = compare(node, $n$, $\bS$, $I$, $\bv$)}
\begin{algorithmic}[1]

    \IF{Leaf node.}
    \label{check2}
       \STATE $\bs_c = $ vector from the leaf node.

       \STATE $I_c = \bs_c^\dag\bS \bS^\dag\bs_c$
       
       \IF {$I_c > I$}
           \STATE $I = I_c$
           \STATE $\bm{s} = \bm{s}_c$
           \RETURN $\bm{s}$
       \ENDIF
   \ENDIF

   \STATE $\bp_c^{(n)} = [0 \quad 0 \ \ldots \ p_{c,n}\ \ldots \ 0
     \quad 0]^\dag$ where $\bp_c$ is the vector of the current node.

   \STATE $I_p = {\bp_c^{(n)}}^\dag \bS \bS^\dag \bp_c^{(n)}$
   \IF{$I_p > I$}
      \IF{$n > N$}
         \STATE $n = 1$
      \ENDIF

      \IF{${\bp_L^{(n)}}^\dag \bS \bS^\dag \bp_L^{(n)} <
        {\bp_R^{(n)}}^\dag \bS \bS^\dag \bp_R^{(n)}$}

        \STATE $n = n + 1$
        \STATE compare(left-child node, $n$, $\bS$, $I$, $\bs$)
      \ELSE
        \STATE $n = n + 1$
        \STATE compare(right-child node, $n$, $\bS$, $I$, $\bs$)
      \ENDIF
   \ELSE
      \STATE $n = n + 1$
      \STATE compare(left-child node, $n$, $\bS$, $I$, $\bs$)
      \STATE compare(right-child node, $n$, $\bS$, $I$, $\bs$)
   \ENDIF
\end{algorithmic}
\label{al10}
\end{algorithm}

This modified kd-tree search does not produce the optimal vector that
maximizes or minimizes the quadratic form, but can perform close to
the optimum when $N$ is large as shown in numerical examples.
Analyzing performance of this scheme is not tractable.  However, the
performance is upper bounded by that of the full search, which was
analyzed in a large system limit by~\cite{dai09, mimo09}.

\section{Numerical Results}
\label{num_re} 

\begin{figure}[h]
\centering
\includegraphics[width=3.25in]{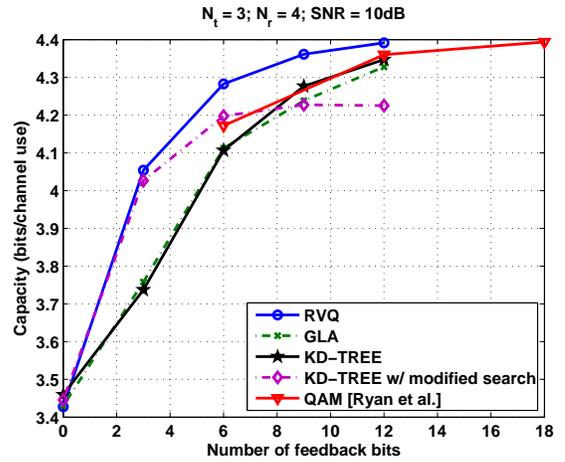}
\caption{Shown is the ergodic capacity of the quantized beamforming vector for
  MIMO channel with 3 transmit and 4 receive antennas.  The performance of
  different quantization schemes are shown with $B$ feedback bits.}
\label{fig1}
\end{figure}

Fig.~\ref{fig1} shows the channel capacity of the quantized beamformer
for the MIMO channel with $N_t = 3$ and $N_r = 4$ and $\rho = 10
\ \text{dB}$.  We compare different quantization schemes and note that
RVQ with full search gives the maximum capacity for a given feedback.
When there is no feedback ($B = 0$), the transmitter deploys a random
beamformer and thus, all schemes give the same performance.  As
available feedback increases, the performance of the quantized
beamforming vector increases with different rates depending on which
quantization scheme is used.  We remark that the proposed kd-tree with
a modified search performs close to RVQ for small $B$.  We also plot
the performance of the GLA and kd-tree schemes from
Section~\ref{tsrvqcb} and also that of the suboptimal QAM codebook
proposed by~\cite{ryan09}.  Reference~\cite{ryan09} also proposed the
optimal QAM codebook scheme that searches for the codebook entry
closest in angle to the optimal eigenvector.  Here we opt to compare
our proposed schemes with the suboptimal QAM codebook since the
suboptimal QAM codebook performs very close to the optimal QAM
codebook with significantly less complexity.  In Fig.~\ref{fig1}, we
see that quantization of the eigenvector does not perform well for
small $B$, but its performance is very close to that of RVQ for large
$B$.  The QAM codebook performs a bit better than the other two
schemes, but is not applicable for $B < 2 N_t$.

\begin{figure}[h]
\centering
\includegraphics[width=3.25in]{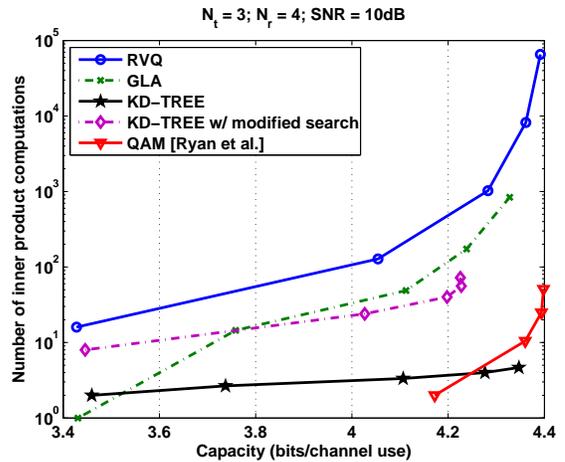}
\caption{The average number of matrix inner product computations
  required to locate the selected entry for various quantization
  codebooks is shown with the ergodic capacity for MIMO channel with $N_t
  = 3$, $N_r = 4$, and $\rho = 10 \ \text{dB}$.}
\label{fig2}
\end{figure}

In addition to channel capacity, we also examine the computational
complexity of each quantization scheme required to search for the
selected entry in the codebook.  Fig.~\ref{fig2} shows the average
number of matrix inner product computations used for each scheme
versus the capacity with the same set of parameters as the previous
figure.  As expected, the RVQ codebook with a full search requires the most
number of inner product computations, which grows exponentially as $B$
increases.  At a capacity of 4.2 bits per channel use, the kd-tree requires
almost two orders of magnitude less computations than RVQ does while the kd-tree with the modified search requires almost one order of magnitude
less.  The QAM codebook is the least complex for a large capacity,
however it is not available for a low capacity.

\begin{figure}[h]
\centering
\includegraphics[width=3.25in]{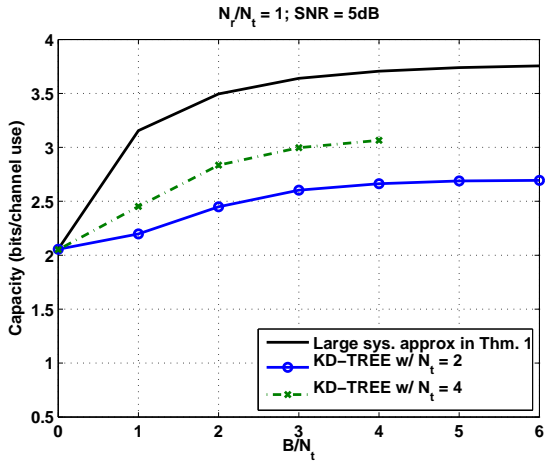}
\caption{The asymptotic capacity approximation is shown with the
  capacity of $2 \times 2$ and $4 \times 4$ channels.}
\label{fig3}
\end{figure}

We compare the large system capacity approximation in Theorem~\ref{t1}
with the capacity of a finite-size MIMO channel with $N_r/N_t = 1$ and
show the comparison in Fig.~\ref{fig3}.  We note that there is a
substantial gap between the theoretical approximation and the simulation
results especially for the $2 \times 2$ system.  The gap narrows as the 
system size increases to $4 \times 4$ and is expected to narrow
further as the size increases.  For the $2 \times 2$ channel, one
feedback bit per transmit antenna does not improve the capacity by much.
However, as the system size increases, one feedback bit per antenna could
potentially increase the capacity as much as 50\% from the
zero-feedback capacity.

\begin{figure}[h]
\centering
\includegraphics[width=3.25in]{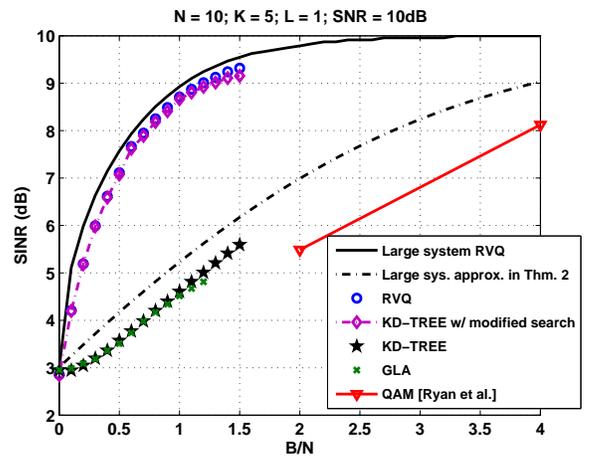}
\caption{SINR with quantized signatures from different quantization
  schemes are compared.}
\label{fig4}
\end{figure}

The simulation results for CDMA are shown in Fig.~\ref{fig4}.  The kd-tree
codebook with a modified search has performance that is close to that of the RVQ codebook while the
eigenvector-quantized schemes perform much worse, especially for small
$\B$.  We also plot the large system approximation in
Theorem~\ref{thm_cdma}.  Unlike the MIMO results, the derived
approximation predicts the performance of the finite-size CDMA quite
well since the system size here ($N = 10$) is much larger than that in
MIMO.  We note that results for the RVQ codebook with large $\B$ are
missing due to the very large memory requirement to store the codebook and
the large computing power required to locate the codebook entry.  In the
case of the GLA and the kd-tree codebooks, the results for large $B$ are also
missing because building a tree with large number of entries also
demands very high computing power.

\begin{figure}[h]
\centering
\includegraphics[width=3.25in]{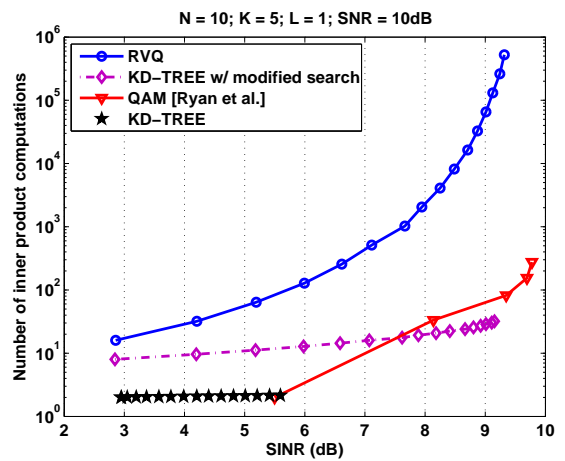}
\caption{The average number of inner product computations to search for
  selected entry in the codebook for each quantization scheme is shown
  with SINR in dB.}
\label{fig5}
\end{figure}

We also show the large system performance of the RVQ codebook derived
in~\cite{dai09}.  We remark that with one feedback bit per processing
gain, SINR for the RVQ codebook is about 4 dB higher than that for the
eigenvector-quantized codebooks.  However, this advantage comes with a
much larger search complexity as we will see in Fig.~\ref{fig5}.

Besides capacity, we would like to compare the complexity of each scheme
to locate the selected entry from the RVQ codebook.  The computational
complexity will be measured by the number of inner products between two
$N$-dimensional vectors.  For an exhaustive search, the number of
inner-product computations increases exponentially with $B$.  For the
proposed binary tree with the nearest-neighbor search, the number of
inner-product computations depends on the tree's depth and only
increases linearly with $B$.  The complexity of the kd-tree search is even
less since in each search step, only one dimension of an $N$-dimensional
codebook entry is required.  Thus, the search complexity of the kd-tree is
proportional to $B/N$.  This also applies to the modified kd-tree
search as well.

We compare the search complexity associated with different codebooks
whose SINR performance is shown in Fig.~\ref{fig5}.  Similar to the
results for the MIMO channel, the average number of inner product
computations used to search the selected entry is shown.  The RVQ
codebook is the most complex while the kd-tree and QAM codebooks are
the least complex.  At 9 dB, the kd-tree codebook with a modified
search requires about 3 orders of magnitude in search complexity less
than RVQ does. The QAM codebook also requires few computations,
however, it is not valid for small number of feedback bits.

\section{Conclusions}
\label{conclude}

We have proposed tree-structured RVQ codebooks to quantize the
transmitted signature for CDMA and the beamforming vector for the MIMO
channel.  The tree structure is generated by either the GLA or kd-tree
algorithms.  For small-to-moderate feedback, the kd-tree codebook with
a modified search has performance that is close to that of the RVQ
with the exhaustive search.  Thus, our proposed scheme complements the
QAM codebook proposed by Ryan {\em et al}~\cite{ryan09}, which does
not perform well in a low-feedback regime.  For large feedback,
quantizing the eigenvector with the GLA or the kd-tree codebooks produces a
performance that is close to RVQ.  The search complexity of the
proposed schemes are a few orders of magnitude less than that of RVQ
for a given performance.  We also approximated the performance with
large system limit, and numerical examples have shown that the
approximations can predict the simulation results well for a
moderate-size system.

In this study, we only consider quantization of a rank-one transmit
beamforming vector in MIMO channels and single-user signature
quantization in CDMA. To extend the existing results to MIMO
channels with an arbitrary-rank precoding matrix or uplink CDMA with
multiple-signature quantization, we can vectorize the precoding matrix or
multiple signatures and apply the GLA or kd-tree algorithms to construct a
tree.  However, the mentioned scheme can be highly suboptimal.  Thus,
improving the scheme and analyzing the associated performance are
interesting open problems.

\bibliographystyle{IEEEtran}
\bibliography{IEEEabrv,tskd}





\end{document}